\documentclass[12pt,prd,showkeys,amsmath,amssymb,nofootinbib]{revtex4-1}
\usepackage{amsmath,amsfonts,amsthm,amscd,amssymb,latexsym}
\usepackage{bm}
\usepackage{dcolumn}
\usepackage{graphicx}
\usepackage{epstopdf}
\usepackage{color}
\usepackage{epsf}
\usepackage{epsfig}
\usepackage{graphicx, epic, eepic, color}
\usepackage[colorlinks=true,urlcolor=blue,linkcolor=blue,citecolor=blue]{hyperref}

\begin{document}

\title{Inertia}

\author{Bahram \surname{Mashhoon}$^{1,2,3}$}
\email{mashhoonb@missouri.edu}

\affiliation{$^1$School of Astronomy, Institute for Research in Fundamental Sciences (IPM), Tehran 19395-5531, Iran\\
$^2$Department of Physics, Sharif University of Technology, Tehran 11155-9161, Iran\\
$^3$Department of Physics and Astronomy, University of Missouri, Columbia, Missouri 65211, USA
}

\date{\today}

\begin{abstract}
Inertia of a particle is due to its mass as well as intrinsic spin. The latter is revealed via the coupling of intrinsic spin with rotation. The spin-rotation coupling and the concomitant spin-gravity coupling are discussed in connection with the nature of inertia. 
The spin-rotation-gravity coupling leads to a gravitomagnetic Stern-Gerlach type of force on the particle that is independent of the particle's mass and thus violates the universality of free fall. This effect is extremely small and its measurement is beyond present capabilities. 
 
\end{abstract}


\maketitle

\section{Introduction}

Inertia is the natural tendency of matter to preserve its state of motion. Inertial physics originated with Newton's formulation of the basic laws of motion~\cite{Cohen}. Newton's laws of motion are valid in global inertial reference frames that are related to each other by Galilean transformations. With the advent of the modern theory of relativity, Galilean invariance was replaced by Lorentz invariance in Minkowski spacetime and later the global inertial frames of reference were replaced by the local inertial frames of reference in the curved spacetime of the general theory of relativity. The Newtonian inertial physics has provided the proper arena for the formulation of the fundamental laws of microphysics. That is, the field equations of Maxwell's electrodynamics and the basic principles of the quantum theory refer to the observations of hypothetical ideal inertial observers; each such observer is free, forever spatially at rest in an inertial frame of reference, and devoid of all limitations associated with actual observers. 

The intrinsic state of a classical particle in Newtonian mechanics is characterized by its mass, $m$, while its extrinsic state in space and time is characterized by its instantaneous position and velocity. On the other hand, the state of a particle in the quantum theory is characterized by its mass and intrinsic spin, which specify the irreducible unitary representations of the inhomogeneous Lorentz group~\cite{Wigner}. Therefore, mass and intrinsic spin determine the inertial properties of a particle in the quantum theory. 

In Newtonian mechanics, the momentum of a particle $\bm{p}$ is its mass multiplied by its velocity with respect to an inertial frame of reference, i.e., $\bm{p} = m \bm{v}$. The principle of inertia, which is Newton's first law of motion, states that momentum $\bm{p}$ is conserved for a free particle.  For a system of particles, the total momentum is the sum of the momenta of the particles.  Newton's third law of motion implies that the net momentum of an isolated two-particle system is conserved and this result can be used, in principle, to define mass within the framework of Newtonian mechanics~\cite{Mach}. 

Inertial forces appear in a reference system that accelerates with respect to an inertial frame of reference.  An inertial force acting on a particle is proportional to its mass.  In  translationally accelerated systems, we find d'Alembert inertial forces and in rotationally accelerated systems, we can observe Coriolis,  centrifugal and Euler inertial forces. Newton's rotating bucket experiment involved the observation of inertial forces and established the reality of inertial frames of reference~\cite{Mashhoon:1984fj, BM1, BM2, BM3, BM4, Mashhoon:1999nr, Lichtenegger:2004re, Mashhoon:2015nea}.  

The inertial properties of mass are well known in classical physics as well as in quantum mechanics~\cite{Werner:1979gi, Staudenmann:1980uqe, Werner, Rauch:2015jkh}; therefore, we turn our attention to the inertial properties of intrinsic spin.  

\section{Inertia of Intrinsic Spin}

Let us consider a thought experiment in which a macroscopic object is set into rotation. Do the intrinsic spins of the particles that constitute the object such as the electrons, protons and neutrons participate in this rotation? We assume that to a hypothetical 
noninertial observer embedded with the object, the intrinsic spins respond to the rotation of the object by rotating in the opposite sense. That is, among other interactions, the intrinsic spin of a particle $\bm{\sigma}$ tends to preserve its direction relative to the local inertial frame of reference and thus appears to the embedded rotating observer to precess in the contrary direction in accordance with the relation
\begin{equation}\label{1}
\frac{d\sigma_i}{dt} + \epsilon_{ijk} \Omega^j \sigma^k = 0\,,  
\end{equation}
where $\bm{\Omega}$ is the angular velocity of rotation of the macroscopic object relative to the local inertial frame of reference. Here, Latin indices $i, j, k,\dots = 1,2,3$. The hypothetical rotating observer embedded with the object would associate a Hamiltonian with this inertial motion of the intrinsic spin given by
\begin{equation}\label{2}
\mathcal{H}_{\rm SR} = - \bm{\sigma} \cdot \bm{\Omega}\, 
\end{equation}
such that the equation of motion of the spin operator
\begin{equation}\label{3}
\hbar\,\frac{d\sigma_k}{dt} = i \,[\mathcal{H}_{\rm SR}, \sigma_k] 
\end{equation}
coincides with Eq.~\eqref{1} on the basis of the general principles of the quantum theory. This heuristic argument establishes the connection between the inertia of intrinsic spin and the spin-rotation Hamiltonian~\eqref{2}. The main purpose of the rest of this section is to furnish the physical basis for the spin-rotation coupling. 

\subsection{Spin-Rotation Coupling} 

To clarify the physical status of the spin-rotation Hamiltonian~\eqref{2}, let us imagine an observer rotating with constant angular velocity $\bm{\Omega}$ in an inertial frame of reference such that its velocity of motion is given by
\begin{equation}\label{4}
\bm{v} = \bm{\Omega}\times\bm{r}\,, 
\end{equation}
where $\bm{r} = (x, y, z)$ is its spatial position. The observer measures the energy of an incident particle of energy $E$ and momentum $\bm{p}$. The measured energy is determined in accordance with the \emph{hypothesis of locality} of the theory of relativity by applying Lorentz transformations point by point along the world line of the observer. The result is $E' = \gamma (E - \bm{v} \cdot \bm{p})$, which can be written via Eq.~\eqref{4} as
\begin{equation}\label{5}
E' = \gamma (E - \bm{\Omega} \cdot \bm{L})\,, 
\end{equation}
where $\bm{L} = \bm{r} \times \bm{p}$ is the orbital angular momentum of the particle. On the other hand, adopting the wave picture, one finds
\begin{equation}\label{6}
E' = \gamma (E - \hbar\,\Omega M)\,, 
\end{equation}
where $M$ is the total (orbital plus spin) ``magnetic" quantum number along the axis of rotation. Indeed, $M = 0, \pm 1, \pm2, \dots$, for a scalar or a vector particle, while $M \mp \tfrac{1}{2} = 0, \pm 1, \pm2, \dots$, for a Dirac particle. We can write this result in the WKB approximation as $E' = \gamma (E - \bm{\Omega} \cdot \bm{J})$, where $\bm{J} = \bm{L} + \bm{S}$ is the total angular momentum of the particle. For an incident plane monochromatic electromagnetic wave of frequency $\omega$ and wave vector $\bm{k}$, $\omega = c\, |\bm{k}|$, for instance, $E = \hbar\, \omega$, $\bm{L} =  \bm{r} \times (\hbar \bm{k})$ and $\bm{S} = \pm\hbar\,\bm{k}/|\bm{k}|$ represents the photon spin. Hence, Eq.~\eqref{5} can be written in a more general form
\begin{equation}\label{7}
E' = \gamma (E - \bm{\Omega} \cdot \bm{L} - \bm{\Omega} \cdot \bm{S})\,,
\end{equation}
since the generator of rotation in the quantum theory is the total angular momentum~\cite{Mashhoon:1988zz, Mashhoon:1992zz}. The semi-classical Eq.~\eqref{7} exhibits the spin-rotation coupling and takes time dilation into account. Intrinsic spin thus couples to rotation in much the same way as macroscopic spin. These results agree with relativistic quantum theory in appropriate limits~\cite{HehlNi, Soares:1995cj, Ryder, Papini:2001un, Papini:2002cp, Lambiase:2004sm, Obukhov:2016vvk}.  

The coupling of orbital angular momentum with rotation leads via interferometry to the Sagnac effect. Gyroscopes based on the Sagnac effect are routinely employed in inertial navigation systems. Similarly, the coupling of intrinsic spin with rotation could be employed, in principle, in the design of new types of gyroscopes~\cite{Fedderke:2024ncj, Mashhoon:2024wvp}.

It is interesting to discuss a simple consequence of Eq.~\eqref{7} involving observers that are all spatially at rest along the $z$ axis and refer their measurements to horizontal axes that rotate uniformly with angular speed $\Omega > 0$ in the positive sense about the $z$ axis. These observers measure the frequency of an incident monochromatic plane electromagnetic wave of definite helicity, frequency $\omega$ and wave vector $(\omega/c) \hat{\bm{z}}$, where $\hat{\bm{z}}$ is a unit vector in the $z$ direction. It follows from Eq.~\eqref{7} that the frequency  of  the radiation as measured by the noninertial observers all along the propagation direction of the wave is
\begin{equation}\label{8}
\omega' = \omega \mp \Omega\,, 
\end{equation}
where the upper (lower) sign refers to radiation of positive (negative) helicity~\cite{Mashhoon:2002fq, Hauck:2003gy}. 

For positive (negative) helicity radiation, the electric and magnetic fields rotate with angular speed $\omega$ in the positive (negative) sense about the $z$ axis. For observers rotating in the positive sense with angular speed $\Omega$, the radiation fields in the positive (negative) helicity case appear to rotate with angular speed $\omega - \Omega$ ( $\omega + \Omega$ ) in the positive (negative) sense about the $z$ axis. This specific combination of angular velocities in the present case is reminiscent of the combination of linear velocities in the Doppler effect; hence the helicity-rotation effect is known in this case as the ``rotational Doppler effect"~\cite{Garetz, E+E, Mashhoon:2024qwj}. 

If the observer rotates uniformly about the $z$ axis on a circular orbit in the $(x, y)$ plane, we find $\omega' = \gamma (\omega \mp \Omega)$, where $\gamma$ is the Lorentz factor. In our present simple treatment, the radiation is incident along the axis of rotation $z$; for  oblique incidence, see~\cite{Mashhoon:2002fq, Hauck:2003gy, Mashhoon:2024qwj} and the references therein. 

Let us note the difference between Eq.~\eqref{8} and the consequence of the transverse Doppler effect in this case, namely, 
\begin{equation}\label{9}
\omega_{\rm D} = \omega\,. 
\end{equation}
This is the result of the standard theory of relativity based on the hypothesis of locality and the invariance of the phase of the radiation under Lorentz transformation. We have 
\begin{equation}\label{10}
\omega' = \omega_{\rm D} \left( 1 \mp \frac{\Omega}{\omega}\right)\,,
\end{equation}
where $\Omega / \omega \to 0$ in the geometric optics limit~\cite{Mashhoon:2024qwj}. The rotational Doppler effect has extensive observational support; indeed, it has been verified for $\omega \gg \Omega$ through the Global Positioning System (GPS), where it is known as the \emph{phase wrap-up}~\cite{Ashby}.

To extend the theory of relativity beyond the WKB limit has been the aim of the nonlocal special relativity theory~\cite{Mashhoon:1997qc, BM5, Mashhoon:2008vr}. A comprehensive treatment of spin-rotation coupling and a more complete list of references are contained in~\cite{BMB}, while the results of recent spin-rotation experiments in connection with neutron physics can be found in~\cite{DSH, DDSH, DDKWLSH, Geerits:2024jdt}.

\subsection{Energy Shift via Rotating Spin Flipper}

The inertial properties of intrinsic spin are completely independent of the inertial properties of mass.  In this connection, it is interesting to consider a situation that can be elucidated equally well using either inertia of mass or inertia of spin. Specifically, we focus on the energy shift that occurs when a spinning particle passes through a rotating spin flipper~\cite{Mashhoon:1998dm, Mashhoon:2005ms}. The inertia of intrinsic spin and the spin-rotation coupling are universal phenomena; however, to simplify matters we consider photons of definite helicity that propagate along the $z$ axis and are normally incidents on a half-wave plate (HWP) in the $(x, y)$ plane that rotates uniformly in the positive sense with angular speed $\Omega > 0$ about the $z$ axis. We employ the nonrelativistic approximation here; for instance, we ignore the deviation of $\gamma$ from unity. The present considerations are generally consistent with the thermodynamics of rotating bodies~\cite{L+L} and dynamics of fluids with spin~\cite{Obukhov:1987yu, ObPi}. The angular momentum of the HWP about the $z$ axis is $\mathcal{J} = \mathcal{I}\, \Omega$ and its kinetic energy is $\mathcal{E} = \tfrac{1}{2}\, \mathcal{I}\, \Omega^2$, where $\mathcal{I}$ is its moment of inertia. Therefore, 
\begin{equation}\label{11}
\left(\frac{\partial \mathcal{E}}{\partial \mathcal{J}}\right)_{\mathcal{I}} = \Omega\,; 
\end{equation}
that is, a small variation in angular momentum $\delta \mathcal{J}$ leads to a small change in energy $\delta \mathcal{E} = \Omega\,\delta \mathcal{J}$ if the moment of inertia is held constant. Imagine a photon of positive helicity and energy $E_{\rm in}$ that is normally incident on the HWP and emerges with negative helicity and energy $E_{\rm out}$; in this case, the photon has deposited angular momentum and energy in the HWP, namely, $\delta \mathcal{J} = 2 \,\hbar$ and $\delta \mathcal{E} = 2 \,\hbar\,\Omega$. Conservation of energy then implies~\cite{Allen, G+A, SKS, GN} 
\begin{equation}\label{12}
E_{\rm out} - E_{\rm in} = - 2 \,\hbar\,\Omega\,. 
\end{equation}

Next, we turn to the inertia of intrinsic spin and consider hypothetical noninertial observers that are embedded with the rotating HWP. According to these observers, the frequency of the incident positive-helicity photon as it enters the HWP is given via spin-rotation coupling by
\begin{equation}\label{13}
\omega' = \omega_{\rm in} - \Omega\,.
\end{equation}
The HWP rotates uniformly; hence, the spacetime inside the plate is stationary and the frequency of the photon inside the HWP as measured by the embedded rotating observers remains constant and equal to  $\omega'$ as a consequence of the invariance under time translation. As the photon exits the HWP, the spin-rotation coupling for the outgoing negative-helicity photon implies 
\begin{equation}\label{14}
\omega' = \omega_{\rm out} + \Omega\,.
\end{equation}
Therefore, 
\begin{equation}\label{15}
\omega_{\rm out} - \omega_{\rm in} = -2\, \Omega\,,
\end{equation}
which, after multiplication by $\hbar$ and noting that $E_{\rm in} = \hbar \,\omega_{\rm in}$, etc., becomes the same as Eq.~\eqref{12}. Inertia of mass is thus compatible with the inertia of intrinsic spin.

\subsection{Spin-Vorticity Coupling}

In a material medium, the spin-rotation coupling can assume a more general form as spin-vorticity coupling. From Eq.~\eqref{4}, namely, $\bm{v} = \bm{\Omega}\times\bm{r}$ and the definition of vorticity $\bm{\mathcal{V}}$,
\begin{equation}\label{16}
\bm{\mathcal{V}} = \bm{\nabla} \times \bm{v}\,,
\end{equation}
we find 
\begin{equation}\label{17}
\bm{\mathcal{V}} = 2\, \bm{\Omega}\,.
\end{equation}
Therefore, the Hamiltonian for the spin-vorticity coupling is
\begin{equation}\label{18}
\mathcal{H}_{\rm SV} = - \frac{1}{2}\, \bm{S} \cdot \bm{\mathcal{V}}\,. 
\end{equation}

In a continuous medium, the vorticity could have gradients which could result in a type of Stern-Gerlach effect and its associated force, namely, 
\begin{equation}\label{18a}
\bm{f} = - \bm{\nabla}{\mathcal{H}_{\rm SV}} =  \frac{1}{2}\, S_k \bm{\nabla} \mathcal{V}^k\,. 
\end{equation}
In this connection, interesting new effects have been explored in spintronics; see~\cite{Mashhoon:2021qtc, Yu:2022vjn, Nozaki} and the references cited therein.

\section{Spin-gravity coupling}

What are the implications of the inertia of intrinsic spin for the gravitational interaction?  Einstein's theory of gravitation  is in good agreement with observational data in the Solar System as well as binary pulsars. Indeed, we are interested in the weak-field approximation of general relativity (GR). Linearized GR leads, among other things, to linear gravitational waves. Gravitational radiation has been detected in ground-based laboratories since 2015 and the results are consistent
 with GR~\cite{GW0, GW}. Moreover, in the weak-field approximation, GR can also describe the linear post-Newtonian gravitational fields of normal astronomical sources such as the Sun, the Earth, etc. This approximate form of GR can be cast in a form that is reminiscent of Maxwell's electrodynamics. That is, the analogy  between Newton's law of gravitation and Coulomb's law of electricity leads to the gravitoelectric description of Newtonian gravitation. Furthermore, magnetic fields are generated by electric currents;  similarly, mass currents generate non-Newtonian gravitomagnetic fields. The Gravity Probe B (GP-B) experiment has measured the exterior gravitomagnetic field of the Earth~\cite{Everitt:2011hp, Everitt:2015qri}. The weak gravitational effects of rotating astronomical masses can be studied using the framework of gravitoelectromagnetism (GEM)~\cite{Mashhoon:2003ax, Mashhoon:2000he, Bini:2021gdb}. 
 
 We consider the weak gravitational field of a  rotating astronomical source of constant mass $\mathfrak{M}$ and angular momentum $\bm{\mathfrak{J}}$. The spacetime metric is given by $g_{\mu\nu} = \eta_{\mu\nu} + h_{\mu\nu}(x)$, where $x^\mu = (ct,\bm{x})$ are spacetime coordinates,  $h_{\mu\nu}(x)$ is the perturbing field that is treated to linear order and $\eta_{\mu\nu} = {\rm diag}(-1, 1, 1, 1)$ is the Minkowski metric tensor of the background spacetime. Here, Greek indices run from $0$ to $3$ and the signature of the spacetime metric is $+2$ in our convention. Neglecting $O(c^{-4})$ terms in the gravitational potentials, the spacetime metric in the weak-field slow-motion approximation  has the GEM form
\begin{equation}\label{19} 
ds^2 = -\,c^2\left(1+2\frac{\Phi_g}{c^2}\right)dt^2-\frac{4}{c}(\bm{A}_g\cdot d\bm{x})dt+\left(1-2\frac{\Phi_g}{c^2}\right) \delta_{ij}dx^idx^j\,,
\end{equation}
where $\Phi_g$ and $\bm{A}_g$ are gravitational potentials given by
\begin{equation}\label{20} 
\Phi_g (t, \bm{x}) = -\,G\int\frac{\rho(ct-|\bm{x}-\bm{x}'|,\bm{x}')}{|\bm{x}-\bm{x}'|}\,d^3x'\,
\end{equation}
and
\begin{equation}\label{20a} 
 \bm{A}_g (t, \bm{x}) = \frac{2G}{c}\int\frac{\bm{j}(ct-|\bm{x}-\bm{x}'|,\bm{x}')}{|\bm{x}-\bm{x}'|}\,d^3x'\,.
\end{equation}
Here, $\rho$ is the mass density and $\bm{j} = \rho \,\bm{v}$ is the mass current.  
 
Henceforth, we confine our attention to stationary systems such that $\rho$ and $\bm{j}$ are independent of time; then,  $\Phi_g (\bm{x})$ reduces to the Newtonian gravitational potential of the system. Moreover, far from the source, we have 
\begin{equation}\label{21}
 \Phi_g \sim \frac{G\, \mathfrak{M}}{|\bm{x}|}\,,\qquad \bm{A}_g \sim \frac{G}{c} \,\frac{\bm{\mathfrak{J}}\times \bm{x}}{|\bm{x}|^3}\,.
\end{equation} 
 
It is useful to define GEM fields  in analogy with electrodynamics via
\begin{equation}\label{22} 
\bm{E}_g = \bm{\nabla}\Phi_g\,, \qquad \bm{B}_g = \bm{\nabla} \times \bm{A}_g\,,
\end{equation}
so that for the exterior of the astronomical source, 
\begin{equation}\label{23}
\bm{E}_g \sim \frac{G\,\mathfrak{M}\,\bm{x}}{|\bm{x}|^3}\,, \qquad  \bm{B}_g \sim \frac{G}{c\,|\bm{x}|^5}\, [\,3 (\bm{\mathfrak{J}} \cdot \bm{x})\, \bm{x} - \bm{\mathfrak{J}}\,|\bm{x}|^2\,]\,.
\end{equation}
The divergence and curl of the GEM fields vanish in the case under consideration here. 

In electrodynamics, Larmor's theorem establishes a local equivalence between the magnetic field and a rotating frame of reference~\cite{Larmor}; similarly, the gravitational Larmor theorem states that the gravitomagnetic field is locally equivalent to a rotation with angular velocity~\cite{Bahram}
\begin{equation}\label{24}
\bm{\Omega}_{\rm L} = -\,{\frac{1}{c}}\,\bm{B}_g\,.
\end{equation}
Here, the equivalence is universal due to the universality of the gravitational interaction; indeed, the gravitational Larmor theorem is essentially the rotational aspect of Einstein's heuristic principle of equivalence~\cite{Mashhoon:2003ax}. It follows from Eq.~\eqref{24} and the spin-rotation Hamiltonian that the intrinsic spin couples to the gravitomagnetic field in accordance with the Hamiltonian
\begin{equation}\label{25}
\mathcal{H}_{\rm SG} = \frac{1}{c} \,\bm{S} \cdot \bm{B}_g\,.
\end{equation}
The intrinsic spin-gravity coupling has been the subject of numerous investigations; see~\cite{DeOT, BM0, BM, Bah1, Bah2, Ramos:2006sb, Mashhoon:2008si, Papini:2007gx, Mashhoon:2013jaa, Tarallo:2014oaa, Fadeev:2020gjk, Mashhoon:2023idh, Vergeles:2022mqu} and the references cited therein.  A comprehensive account of spin-gravity interaction is contained in~\cite{Lambiase:2021txu}.

In ground-based laboratory experiments, the spin-rotation-gravity coupling for a  spin-$\tfrac{1}{2}$ particle with intrinsic spin $\bm{\sigma}$ leads to the Hamiltonian $-\bm{\sigma}\cdot (\bm{\Omega} - \bm{B}_g/c)_{\oplus}$, where the magnitude of the energy difference for the spin polarized vertically up and down is $\hbar \Omega_{\oplus} \approx 5 \times 10^{-20}$ eV for the coupling to Earth's rotation, while $\hbar |\bm{B}_g|/c  \approx 10^{-29}$ eV for the coupling to the gravitomagnetic field on the surface of the Earth. The measurement of the spin-gravity coupling is, in general,  a few orders of magnitude beyond current capabilities. 

\section{Gravitomagnetic Stern-Gerlach Effect}

In the linear post-Newtonian framework of the GEM, the dependence of $\bm{B}_g(\bm{x})$ on position implies that there exists a gravitomagnetic Stern-Gerlach force
\begin{equation}\label{26}
-\frac{1}{c} \,\bm{\nabla}(\bm{S} \cdot \bm{B}_g) = -\frac{1}{c} \,(\bm{S} \cdot \bm{\nabla})\,\bm{B}_g\,.
\end{equation}
Employing the gravitomagnetic field given in Eq.~\eqref{23}, this gravitomagnetic Stern-Gerlach type of force has components $\mathbb{F}^k$, $k = 1, 2, 3$,  given by
\begin{equation}\label{27}
\mathbb{F}^k \sim  - \frac{3 G}{c^2\,|\bm{x}|^7}\,\left \{[(\bm{\mathfrak{J}} \cdot \bm{x}) S^k + (\bm{S} \cdot \bm{\mathfrak{J}})x^k + (\bm{S} \cdot  \bm{x})\mathfrak{J}^k] |\bm{x}|^2 - 5 (\bm{S} \cdot  \bm{x})(\bm{\mathfrak{J}} \cdot \bm{x}) x^k\right \}\,,
\end{equation}
which are independent of the particle's mass and thus violate the universality of free fall. In the correspondence limit, the gravitomagnetic Stern-Gerlach force is in agreement with the classical Mathisson spin-curvature force~\cite{Mashhoon:2021qtc}. 

Imagine a spin-$\tfrac{1}{2}$ particle of mass $m$ and intrinsic spin $\bm{\sigma}$ that is held at rest in the exterior gravitational field of a rotating source given by Eq.~\eqref{19}. The gravitational force on the particle is
\begin{equation}\label{28}
\bm{\mathcal{F}} = - m \bm{E}_g - \frac{1}{c} \,(\bm{\sigma} \cdot \bm{\nabla})\bm{B}_g\,,
\end{equation}
so that the mass of the particle couples to the gravitoelectric field and its spin couples to the gravitomagnetic field. 

\section{Free Fall is Not Universal} 

The dependence of the gravitational force upon the intrinsic spin of the particle violates the universality of free fall. Let us compute the weight of the spin-$\tfrac{1}{2}$ particle that is spatially at rest using Eqs.~\eqref{27} and~\eqref{28}, namely, 
\begin{equation}\label{29}
w(\bm{x}) := - \bm{\mathcal{F}}\cdot \frac{\bm{x}}{|\bm{x}|} =  m g - \frac{3}{c |\bm{x}|} \bm{\sigma} \cdot \bm{B}_g\,,
\end{equation}
where $g = G \mathfrak{M}/|\bm{x}|^2$ is the Newtonian acceleration of gravity. In a ground-based laboratory where the vertical direction is chosen to be the axis of quantization, the weight of the particle is given by
\begin{equation}\label{30}
w_{\pm}  =  m g_{\oplus} \mp 3\, \frac{\hbar G \mathfrak{J}_{\oplus}}{c^2 R_{\oplus}^4} \sin \vartheta\,,
\end{equation}
where $R_{\oplus}$ is the radius of the Earth and $\vartheta$ is the geographical latitude measured from the equator. Here, the upper (lower) sign indicates that the spin of the particle is polarized vertically up (down). Writing Eq.~\eqref{30} as  
\begin{equation}\label{31}
w_{\pm}  =  m g_{\oplus} (1 \mp \epsilon_{\oplus} \sin \vartheta)\,,
\end{equation}
we find 
\begin{equation}\label{32}
\epsilon_{\oplus}  \approx \frac{\hbar \Omega_{\oplus}}{m c^2}\,.
\end{equation}
Therefore, for a neutron, say, the inertia of its intrinsic spin leads to a violation of the universality of the gravitational free fall of order
\begin{equation}\label{33}
 \frac{\hbar \Omega_{\oplus}}{m_{\rm n} c^2} \approx \tfrac{1}{2} \times 10^{- 30}\,, 
\end{equation}
which is extremely small and its measurement is beyond current capabilities by many orders of magnitude. Indeed, the universality of free fall has been tested thus far at the level of $10^{-15}$~\cite{MICROSCOPE:2022doy}. 

\section{DISCUSSION}

We have argued that there is a direct coupling between the intrinsic spin of a particle and rotation as a consequence of the inertia of intrinsic spin. Accelerated motion consists in general of translational acceleration as well as rotation. Is there a \emph{direct} coupling between intrinsic spin and translational acceleration as well? A direct interaction Hamiltonian of the form $\bm{S}\cdot \bm{a}/c$ or $\bm{S}\cdot \bm{g}/c$ would violate parity and time-reversal invariance; here, $\bm{a}$ represents the translational acceleration and $\bm{g}$ represents the acceleration of gravity.  The $\bm{S}\cdot \bm{g}/c$ term has been ruled out observationally~\cite{Bah2, W+R}. Furthermore, no such direct spin-acceleration coupling has been found in the study of wave equations in translationally accelerated systems in Minkowski spacetime and in the corresponding gravitational fields~\cite{Bah3, Mashhoon:2004rz, Bini:2004kz}.

Finally, in any experiment involving the gravitational interaction, the whole mass-energy content of the universe must be taken into account~\cite{Bah2}. In this connection, we have assumed that the influence of the gravitomagnetic field of distant rotating masses on local physics can be neglected. It appears that this assumption is consistent with the results of the GP-B experiment~\cite{Everitt:2011hp, Everitt:2015qri}.

 \section*{Acknowledgments}
 
 I am grateful to Yuri Obukhov for helpful discussions. 

\appendix



\begin{thebibliography}{90}

\bibitem{Cohen}
I. B. Cohen, 
\emph{The Birth of a New Physics}
(Doubleday Anchor Books, Garden city, NY, 1960). 

\bibitem{Wigner}
E. P. Wigner, 
``Unitary Representations of the Inhomogeneous Lorentz Group", 
Ann. Math. \textbf{40}, 149-204 (1939). 

\bibitem{Mach}
E. Mach, 
\emph{The Science of Mechanics}
(Open Court, La Salle, 1960).

\bibitem{Mashhoon:1984fj}
B.~Mashhoon, F.~W.~Hehl, and D.~S.~Theiss,
``On the Gravitational effects of rotating masses - The Thirring-Lense Papers",
Gen. Relativ. Gravit. \textbf{16}, 711-750 (1984).

\bibitem{BM1}
B. Mashhoon, 
``Complementarity of Absolute and Relative Motion",
Phys. Lett.  A \textbf{126}, 393-399 (1988).

\bibitem{BM2}
B. Mashhoon,
``Quantum Theory and the Origin of Inertia",
Found. Phys. Lett. \textbf{6}, 545-560 (1993).

\bibitem{BM3}
B. Mashhoon,
``On the Relativity of Rotation",
in: \emph{Directions in General Relativity: Papers in Honor of Dieter Brill}, edited by B. L. Hu and T. A. Jacobson
(Cambridge University Press, Cambridge, 1993), pp. 182-194.

\bibitem{BM4}
B. Mashhoon,
``On the Origin of Inertial Accelerations",
Nuovo Cimento B \textbf{109}, 187-199 (1994).

\bibitem{Mashhoon:1999nr}
B.~Mashhoon, F.~Gronwald, and H.~I.~M.~Lichtenegger,
``Gravitomagnetism and the clock effect",
Lect. Notes Phys. \textbf{562}, 83-108 (2001).
[arXiv:gr-qc/9912027 [gr-qc]]

\bibitem{Lichtenegger:2004re}
H.~Lichtenegger and B.~Mashhoon,
``Mach's principle",
in: \emph{The Measurement of Gravitomagnetism: A Challenging Enterprise}, edited by L. Iorio 
(Nova Science, New York, USA, 2007), pp. 13-25.
[arXiv:physics/0407078 [physics.hist-ph]]


\bibitem{Mashhoon:2015nea}
B.~Mashhoon,
``Mach\textquoteright{}s Principle and the Origin of Inertia",
Fundam. Theor. Phys. \textbf{183}, 177-187 (2016).
[arXiv:1509.01869 [gr-qc]]

\bibitem{Werner:1979gi}
S.~A.~Werner, J.~L.~Staudenmann and R.~Colella,
``Effect of Earth's Rotation on the Quantum Mechanical Phase of the Neutron",
Phys. Rev. Lett. \textbf{42}, 1103-1106 (1979).

\bibitem{Staudenmann:1980uqe}
J.~L.~Staudenmann, S.~A.~Werner, R.~Colella, and A.~W.~Overhauser,
``Gravity and inertia in quantum mechanics",
Phys. Rev. A \textbf{21}, no.5, 1419 (1980).

\bibitem{Werner}
S. A. Werner, 
``Does a neutron know that the earth is rotating?", 
Gen. Relativ. Gravit. \textbf{40}, 921-934 (2008).


\bibitem{Rauch:2015jkh}
H.~Rauch and S.~A.~Werner,
\emph{Neutron Interferometry : Lessons in Experimental Quantum Mechanics}, 2nd edn
(Oxford University Press, Oxford, UK, 2015).

\bibitem{Mashhoon:1988zz}
B.~Mashhoon,
``Neutron interferometry in a rotating frame of reference",
Phys. Rev. Lett. \textbf{61}, 2639-2642 (1988).

\bibitem{Mashhoon:1992zz}
B.~Mashhoon,
``Mashhoon replies",
Phys. Rev. Lett. \textbf{68}, 3812-3812 (1992).

\bibitem{HehlNi}
F. W. Hehl and W.-T. Ni, ``Inertial effects of a Dirac particle'',
Phys. Rev. D {\bf 42}, 2045-2048 (1990).



\bibitem{Soares:1995cj}
I.~D.~Soares and J.~Tiomno,
``The physics of the Sagnac-Mashhoon effects",
Phys. Rev. D \textbf{54}, 2808-2813 (1996).

\bibitem{Ryder}
L. Ryder, 
``Relativistic treatment of inertial spin effects",
J. Phys. A: Math. Gen. \textbf{31},  2465-2469 (1998).


\bibitem{Papini:2001un}
G.~Papini and G.~Lambiase,
``Spin rotation coupling in muon g-2 experiments",
Phys. Lett. A \textbf{294}, 175-178 (2002).
[arXiv:gr-qc/0106066 [gr-qc]]

\bibitem{Papini:2002cp}
G.~Papini,
``Parity and time reversal in the spin-rotation interaction",
Phys. Rev. D \textbf{65}, 077901 (2002).
[arXiv:gr-qc/0201098 [gr-qc]]

\bibitem{Lambiase:2004sm}
G.~Lambiase and G.~Papini,
``Discrete symmetries in the spin-rotation interaction",
Phys. Rev. D \textbf{70}, 097901 (2004).

\bibitem{Obukhov:2016vvk}
Y.~N.~Obukhov, A.~J.~Silenko, and O.~V.~Teryaev,
``Manifestations of the rotation and gravity of the Earth in high-energy physics experiments",
Phys. Rev. D \textbf{94}, no.4, 044019 (2016).
[arXiv:1608.03808 [gr-qc]]

\bibitem{Fedderke:2024ncj}
M.~A.~Fedderke, R.~Harnik, D.~E.~Kaplan, S.~Posen, S.~Rajendran, F.~Serra, and V.~P.~Yakovlev,
``Precision gyroscope from the helicity of light'',
Phys. Rev. A \textbf{111}, 043502 (2025).
[arXiv:2406.16178 [physics.optics]]



\bibitem{Mashhoon:2024wvp}
B.~Mashhoon and Y.~N.~Obukhov,
``Spin-of-light gyroscope and the spin-rotation coupling",
Phys. Rev. D \textbf{110}, no.10, 104015 (2024).
[arXiv:2408.07799 [quant-ph]]


\bibitem{Mashhoon:2002fq}
B.~Mashhoon, ``Modification of the Doppler effect due to the helicity-rotation coupling'',
Phys. Lett. A \textbf{306}, 66-72 (2002).
[arXiv:gr-qc/0209079 [gr-qc]]

  


\bibitem{Hauck:2003gy}
J.~C.~Hauck and B.~Mashhoon, ``Electromagnetic waves in a rotating frame of reference'',
Ann. Phys. (Berlin) {\bf 12}, 275-288 (2003).
 [arXiv:gr-qc/0304069 [gr-qc]]


\bibitem{Garetz}
B. A. Garetz,
 ``Angular Doppler Effect", 
 J. Opt. Soc. Am. \textbf{71},  609-611 (1981). 
 
\bibitem{E+E}
O. Emile and J. Emile,
``Rotational Doppler Effect: A Review",
Ann. Phys. (Berlin) \textbf{535}, 2300250 (2023). 

\bibitem{Mashhoon:2024qwj}
B.~Mashhoon,
``Rotational Doppler Effect and Spin-Rotation Coupling",
[arXiv:2403.17151 [gr-qc]].


\bibitem{Ashby}
N. Ashby, 
``Relativity in the Global Positioning System", 
Living Rev. Relativ. \textbf{6}, 1 (2003). 

\bibitem{Mashhoon:1997qc}
B.~Mashhoon,
``On the spin-rotation-gravity coupling",
Gen. Relativ. Gravit. \textbf{31}, 681-691 (1999).
[arXiv:gr-qc/9803017 [gr-qc]]


\bibitem{BM5}
B. Mashhoon, 
``Nonlocal Electrodynamics of Rotating Systems",
Phys. Rev. A \textbf{72}, 052105 (2005). 
[arXiv: hep-th/0503205]

\bibitem{Mashhoon:2008vr}
B.~Mashhoon,
``Nonlocal Special Relativity",
Ann. Phys. (Berlin) \textbf{520}, 705-727 (2008).
[arXiv:0805.2926 [gr-qc]]

\bibitem{BMB}
B. Mashhoon, 
\emph{Nonlocal Gravity} 
(Oxford University Press, Oxford, UK, 2017).

\bibitem{DSH}
B. Demirel, S. Sponar, and Y. Hasegawa, 
``Measurement of the spin-rotation coupling in neutron polarimetry'', 
New J. Phys. {\bf 17}, 023065 (2015).

\bibitem{DDSH}
A. Danner, B. Demirel, S. Sponar, and Y. Hasegawa, ``Development and performance of
a miniaturised spin rotator suitable for neutron interferometer experiments'',
J. Phys. Commun. {\bf 3}, 035001 (2019).

\bibitem{DDKWLSH}
A. Danner, B. Demirel, W. Kersten, R. Wagner, H. Lemmel, S. Sponar, and Y. Hasegawa, 
``Spin-rotation coupling observed in neutron interferometry'',  
npj Quantum Information {\bf 6}, 23 (2020).
[arXiv:1904.07085 [quant-ph]]


\bibitem{Geerits:2024jdt}
N.~Geerits, S.~Sponar, K.~E.~Steffen, W.~M.~Snow, S.~R.~Parnell, G.~Mauri, G.~N.~Smith, R.~M.~Dalgliesh and V.~de Haan,
``Measuring the angular momentum of a neutron using Earth's rotation",
Phys. Rev. Res. \textbf{7}, no.1, 013046 (2025).
[arXiv:2407.09307 [quant-ph]]



\bibitem{Mashhoon:1998dm}
B.~Mashhoon, R.~Neutze, M.~Hannam, and G.~E.~Stedman,
``Observable frequency shifts via spin-rotation coupling'',
Phys. Lett. A {\bf 249}, 161-166 (1998).
[arXiv:gr-qc/9808077 [gr-qc]]

\bibitem{Mashhoon:2005ms}
B.~Mashhoon and H.~Kaiser,
``Inertia of intrinsic spin",
Physica B \textbf{385}, 1381-1383 (2006).
[arXiv:quant-ph/0508182 [quant-ph]]

\bibitem{L+L}
L. D. Landau and E. M. Lifshitz,
\emph{Statistical Physics}
(Pergamon Press, Oxford, UK, 1969).

\bibitem{Obukhov:1987yu}
Y.~N.~Obukhov and V.~A.~Korotky,
``The Weyssenhoff fluid in Einstein-Cartan theory",
Classical Quantum Gravity \textbf{4}, 1633-1657 (1987).

\bibitem{ObPi}
Y.~N.~Obukhov and O.~B.~Piskareva,
``Spinning fluid in general relativity",
Classical Quantum Gravity \textbf{6}, L15-L19 (1989).



\bibitem{Allen}
P. J. Allen,
 ``A radiation torque experiment", 
 Am. J. Phys. \textbf{34}, 1185-1192 (1966).
 
 \bibitem{G+A}
B. A. Garetz and S. Arnold,
  ``Variable frequency shifting of circularly polarized laser radiation via a rotating half-wave plate",
 Opt. Commun. \textbf{31}, 1-3 (1979).
 
\bibitem{SKS}
R. Simon, H. J. Kimble and E. C. G. Sudarshan,
``Evolving Geometric Phase and Its Dynamical Manifestation as a Frequency Shift: An Optical Experiment",
Phys. Rev. Lett. \textbf{61}, 19-22 (1988).

\bibitem{GN}
G. Nienhuis, 
``Doppler effect induced by rotating lenses", 
Opt. Commun. \textbf{132}, 8-14 (1996).

\bibitem{Mashhoon:2021qtc}
B.~Mashhoon, ``Gravitomagnetic Stern-Gerlach force'',
Entropy {\bf 23}, 445 (2021).
[arXiv:2102.06433 [gr-qc]]


\bibitem{Yu:2022vjn}
T.~Yu, Z.~Luo, and G.~E.~W.~Bauer, ``Chirality as generalized spin-orbit interaction in spintronics'',
Phys. Rept. {\bf 1009}, 1-115 (2023).
 [arXiv:2206.05535 [cond-mat.mes-hall]]

\bibitem{Nozaki}
Y. Nozaki, H. Sukegawa, S. Watanabe, S. Yunoki, T. Horaguchi, H. Nakayama, K. Yamanoi, Z. Wen, C. He, J. Song, T. Ohkubo, S. Mitani, K. Maezawa, D. Nishikawa, S. Fujii, M. Matsuo, J. Fujimoto, and S. Maekawa,  
``Gyro-spintronic material science using vorticity gradient in solids",
Science and Technology of Advanced Materials \textbf{26}, no.1, 2428153 (2025).  

\bibitem{GW0}
B. P. Abbott, R. Abbott, T. D. Abbott, M. R. Abernathy, F. Acernese, K. Ackley, C. Adams, T. Adams, P. Addesso, R. X. Adhikari,  \emph{et al.} (LIGO Scientific Collaboration and Virgo Collaboration),
``Observation of Gravitational Waves from a Binary Black Hole Merger",
Phys. Rev. Lett. \textbf{116}, 061102 (2016).


\bibitem{GW}
B.~P.~Abbott, \emph{et al.} (LIGO Scientific Collaboration and Virgo Collaboration),
``GW170817: Observation of gravitational waves from a binary neutron star inspiral",
Phys. Rev. Lett. \textbf{119}, 161101 (2017).

\bibitem{Everitt:2011hp}
C.~W.~F.~Everitt, D.~B.~DeBra, B.~W.~Parkinson, J.~P.~Turneaure, J.~W.~Conklin, M.~I.~Heifetz, G.~M.~Keiser, A.~S.~Silbergleit, T.~Holmes, and J.~Kolodziejczak, \textit{et al.}
``Gravity Probe B: Final results of a space experiment to test General Relativity'',
Phys. Rev. Lett. \textbf{106}, 221101 (2011).
[arXiv:1105.3456 [gr-qc]]

\bibitem{Everitt:2015qri}
C.~W.~F.~Everitt, B.~Muhlfelder, D.~B.~DeBra, B.~W.~Parkinson, J.~P.~Turneaure, A.~S.~Silbergleit, E.~B.~Acworth, M.~Adams, R.~Adler, and W.~J.~Bencze, \textit{et al.}
``The Gravity Probe B test of general relativity",
Classical Quantum Gravity \textbf{32}, 224001 (2015).



\bibitem{Mashhoon:2003ax} 
 B.  Mashhoon,
 ``Gravitoelectromagnetism: A brief review'',
in: \emph{The Measurement of Gravitomagnetism: A Challenging Enterprise}, L.  Iorio, ed. (Nova Science, New York, USA, 2007), pp. 29-39. 
[arXiv: gr-qc/0311030]
  
\bibitem{Mashhoon:2000he} 
B. Mashhoon, 
``Gravitoelectromagnetism'', in \emph{Reference Frames and Gravitomagnetism}, 
J.-F. Pascual-Sanchez, L. Floria, A. San Miguel, and F. Vicente, eds. 
(World Scientific, Singapore, 2001), pp. 121-132. 
 [arXiv: gr-qc/0011014]
  
\bibitem{Bini:2021gdb}
D.~Bini, B.~Mashhoon, and Yu.~N.~Obukhov, ``Gravitomagnetic helicity'',
Phys. Rev. D {\bf 105}, 064028 (2022).
 [arXiv:2112.07550 [gr-qc]]

\bibitem{Larmor}
J. Larmor, 
``On the theory of the magnetic influence on spectra; and on the radiation from moving ions'', 
Phil. Mag. {\bf 44}, 503-512 (1897).


\bibitem{Bahram}
B. Mashhoon,
``On the gravitational analogue of Larmor's theorem'',
Phys. Lett. A {\bf 173}, 347-354 (1993).

\bibitem{DeOT}
C. G. de Oliveira and J. Tiomno, ``Representations of Dirac equation in general relativity'',
Nuovo Cimento {\bf 24}, 672-687 (1962).

\bibitem{BM0}
B.~Mashhoon,
``Can Einstein's Theory of Gravitation be Tested Beyond the Geometrical Optics Limit?",
Nature \textbf{250}, 316 (1974). 

\bibitem{BM}
B.~Mashhoon, ``Influence of gravitation on the propagation of electromagnetic radiation'', 
Phys. Rev. D {\bf 11}, 2679-2684 (1975).

\bibitem{Bah1}
B. Mashhoon, ``On the coupling of intrinsic spin with the rotation of the Earth'',
Phys. Lett. A {\bf 198}, 9-13 (1995).

\bibitem{Bah2} 
B. Mashhoon, ``Gravitational couplings of intrinsic spin'',
Classical Quantum Gravity {\bf 17}, 2399-2409 (2000).
 [arXiv: gr-qc/0003022]

\bibitem{Ramos:2006sb}
J.~Ramos and B.~Mashhoon,
``Helicity-rotation-gravity coupling for gravitational waves",
Phys. Rev. D \textbf{73}, 084003 (2006).
[arXiv:gr-qc/0601054 [gr-qc]]

\bibitem{Mashhoon:2008si}
B.~Mashhoon,
``Spin-Gravity Coupling",
Acta Phys. Polon. Supp. \textbf{1}, 113-122 (2008).
[arXiv:0801.2134 [gr-qc]]



\bibitem{Papini:2007gx}
G.~Papini,
``Spin-gravity coupling and gravity-induced quantum phases",
Gen. Relativ. Gravit. \textbf{40}, 1117-1144 (2008).
[arXiv:0709.0819 [gr-qc]]


\bibitem{Mashhoon:2013jaa}
B.~Mashhoon and Yu.~N.~Obukhov, ``Spin precession in inertial and gravitational fields'',
Phys. Rev. D {\bf 88}, 064037 (2013).
 [arXiv:1307.5470 [gr-qc]]



\bibitem{Tarallo:2014oaa}
M. G. Tarallo, T. Mazzoni, N. Poli, D. V. Sutyrin, X. Zhang, and G. M. Tino, 
``Test of Einstein Equivalence Principle for 0-spin and half-integer-spin atoms: Search
for spin-gravity coupling effects'',
Phys. Rev. Lett. {\bf 113}, 023005 (2014).
 [arXiv:1403.1161 [physics.atom-ph]]

\bibitem{Fadeev:2020gjk}
P.~Fadeev, T.~Wang, Y.~B.~Band, D.~Budker, P.~W.~Graham, A.~O.~Sushkov, and D.~F.~J.~Kimball,
``Gravity Probe Spin: Prospects for measuring general-relativistic precession of intrinsic
spin using a ferromagnetic gyroscope'',
Phys. Rev. D {\bf 103}, 044056 (2021).
 [arXiv:2006.09334 [gr-qc]]


\bibitem{Mashhoon:2023idh}
B.~Mashhoon, M.~Molaei, and Yu.~N.~Obukhov, ``Spin-Gravity Coupling in a Rotating Universe'',
Symmetry {\bf 15}, 1518 (2023).
 [arXiv:2304.08835 [gr-qc]]


\bibitem{Vergeles:2022mqu}
S.~N.~Vergeles, N.~N.~Nikolaev, Yu.~N.~Obukhov, A.~J.~Silenko, and O.~V.~Teryaev,
``General relativity effects in precision spin experimental tests of fundamental symmetries",
Phys. Usp. {\bf 66}, 109-147 (2023).
 [arXiv:2204.00427 [hep-th]]
 
\bibitem{Lambiase:2021txu}
G.~Lambiase and G.~Papini,
\emph{The Interaction of Spin with Gravity in Particle Physics: Low Energy Quantum Gravity}
(Springer Nature Switzerland AG 2021)
[Lect. Notes Phys. \textbf{993}, 190 pp. (2021)].


\bibitem{MICROSCOPE:2022doy}
P.~Touboul \textit{et al.} [MICROSCOPE],
``MICROSCOPE Mission: Final Results of the Test of the Equivalence Principle",
Phys. Rev. Lett. \textbf{129}, no.12, 121102 (2022).
[arXiv:2209.15487 [gr-qc]]

\bibitem{W+R}
D. J. Wineland and N. F. Ramsey,
``Atomic Deuterium Maser",
Phys. Rev. A \textbf{5}, 821-837 (1972).

\bibitem{Bah3}
B.~Mashhoon,
``Electrodynamics in a  linearly accelerated system",
Phys. Lett. A \textbf{122}, 67-72 (1987).

\bibitem{Mashhoon:2004rz}
B.~Mashhoon,
``Nonlocal electrodynamics of linearly accelerated systems",
Phys. Rev. A \textbf{70}, 062103 (2004).
[arXiv:hep-th/0407278 [hep-th]]

\bibitem{Bini:2004kz}
D.~Bini, C.~Cherubini, and B.~Mashhoon,
``Spin, acceleration and gravity",
Classical Quantum Gravity \textbf{21}, 3893-3908 (2004).
[arXiv:gr-qc/0406061 [gr-qc]]




\end{thebibliography}
\end{document}